\documentclass[prd,aps,nofootinbib,twocolumn,preprintnumbers]{revtex4}
\pdfoutput=1
\RequirePackage{snapshot} 

\usepackage{amsmath}
\usepackage{amssymb}
\usepackage{graphicx}
\usepackage{booktabs}
\usepackage{hyperref}
\usepackage{algorithmicx}
\usepackage{algorithm}
\usepackage{algpseudocode}
\usepackage[caption=false]{subfig}

\AtBeginDocument{
\heavyrulewidth=.08em
\lightrulewidth=.05em
\cmidrulewidth=.03em
\belowrulesep=.65ex
\belowbottomsep=0pt
\aboverulesep=.4ex
\abovetopsep=0pt
\cmidrulesep=\doublerulesep
\cmidrulekern=.5em
\defaultaddspace=.5em
}

\usepackage{xcolor}

\begin{document}

\title{Jet grooming through reinforcement learning}

\preprint{OUTP-19-05P}
\preprint{TIF-UNIMI-2019-2}

\newcommand{\OXaff}{Rudolf Peierls Centre for Theoretical Physics,
  University of Oxford,\\
  Clarendon Laboratory, Parks Road, Oxford OX1 3PU}

\newcommand{\MUaff}{TIF Lab, Dipartimento di Fisica,
  Universit\`a degli Studi di Milano and INFN Milan,\\
  Via Celoria 16, 20133, Milano, Italy}

\author{Stefano Carrazza}
\affiliation{\MUaff}
\author{Fr\'ed\'eric A. Dreyer}
\affiliation{\OXaff}

\begin{abstract}
  We introduce a novel implementation of a reinforcement learning (RL)
  algorithm which is designed to find an optimal jet grooming
  strategy, a critical tool for collider experiments.
  The RL agent is trained with a reward function constructed to
  optimize the resulting jet properties, using both signal and
  background samples in a simultaneous multi-level training.
  We show that the grooming algorithm derived from the deep RL agent
  can match state-of-the-art techniques used at the Large Hadron
  Collider, resulting in improved mass resolution for boosted objects.
  Given a suitable reward function, the agent learns how to train a
  policy which optimally removes soft wide-angle radiation,
  allowing for a modular grooming technique that can be applied in a wide
  range of contexts.
  These results are accessible through the corresponding
  \texttt{GroomRL} framework.
\end{abstract}

\maketitle

\section{Introduction}
\label{sec:introduction}

Jets are one of the most common objects appearing in proton-proton
colliders such as the Large Hadron Collider (LHC) at CERN.
They are defined as collimated bunches of high-energy particles, which
emerge from the interactions of quarks and gluons, the fundamental
constituents of the proton~\cite{Sterman:1977wj,Salam:2009jx}.
In modern analyses, final-state particle momenta are mapped to jet
momenta using a sequential recombination algorithm with a single free
parameter, the jet radius $R$, which defines up to which angle
particles can get recombined into a given
jet~\cite{Ellis:1993tq,Dokshitzer:1997in,Cacciari:2008gp}.

\begin{figure}
  \centering
  \includegraphics[width=0.95\linewidth]{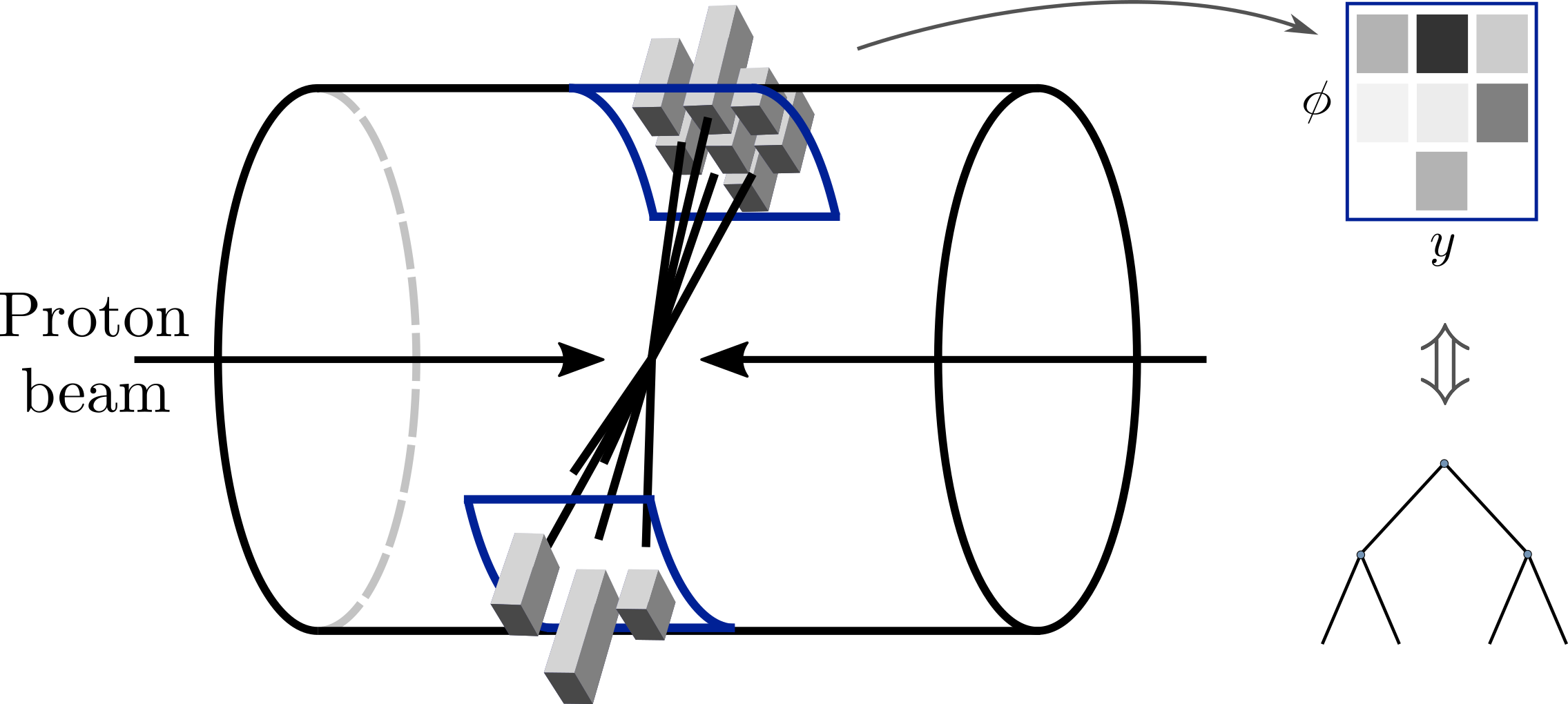}
  \caption{Jets emerging from a proton-proton collision at the LHC,
    and their representation as images in rapidity-azimuth ($y,\phi$) space or as
    clustering trees.}
\label{fig:jet-display}
\end{figure}

An example of an LHC collision resulting in two jets is shown in
figure~\ref{fig:jet-display}, where the towers correspond to energy
deposits in the calorimeter.
The right-hand side gives a schematic visualization of two different
representations of jets, either as an image where the pixel intensity
encodes the energy flow in that phase-space
region~\cite{Cogan:2014oua}, or as a tree defined by the recombination
sequence of the jet algorithm.

Due to the very high energies of its collisions, the LHC is routinely
producing heavy particles, such as top quarks and vector bosons, with
transverse momenta far greater than their rest mass.
When these objects are sufficiently energetic (or boosted), they can
often generate very collimated decays, which are then reconstructed as
a single fat jet.
These fat jets originating from boosted objects can be distinguished
from standard quark and gluon jets by studying differences in their
radiation patterns.
Since the advent of the LHC program, the physics of the substructure
of jets has matured into a remarkably active field of research that
has become notably conducive to applications of recent Machine
Learning
techniques~\cite{Almeida:2015jua,deOliveira:2015xxd,Baldi:2016fql,Guest:2016iqz,Kasieczka:2017nvn,Louppe:2017ipp,deOliveira:2017pjk,Shimmin:2017mfk,Datta:2017rhs,Larkoski:2017jix,Pearkes:2017hku,Butter:2017cot,Komiske:2017ubm,Andreassen:2018apy,Metodiev:2018ftz,Dreyer:2018nbf,Komiske:2018cqr,Martinez:2018fwc,Datta:2019ndh,Kasieczka:2019dbj,Qu:2019gqs}.

A particularly useful set of tools for experimental analyses are jet
grooming
algorithms~\cite{Butterworth:2008iy,Ellis:2009me,Krohn:2009th,Dasgupta:2013ihk,Larkoski:2014wba,Dreyer:2018tjj},
defined as a post-processing treatment of jets to remove soft
wide-angle radiation which is not associated with the underlying hard
substructure.
Grooming techniques play a crucial role in Standard Model
measurements~\cite{Aaboud:2017qwh,Sirunyan:2018xdh} and in improving
the boson- and top-tagging efficiencies at the LHC.

In this article we introduce a novel framework, which we call {\tt GroomRL}, to train a grooming
algorithm using reinforcement learning (RL)~\cite{DBLP:journals/corr/MnihKSGAWR13,silver2017mastering}.
To this end, we decompose the problem of jet grooming into successive
steps for which a reward function can be designed taking into account the 
physical features that characterize such a system.
We then use a modified implementation of a Deep Q-Network (DQN)
agent~\cite{DBLP:journals/corr/MnihKSGAWR13,mnih2015humanlevel} and
train a dense neural network (NN) to optimally remove radiation
unassociated from the core of the jet.
The trained model can then be applied on other data sets, showing
improved resolution compared to state-of-the-art techniques as well as
a strong resilience to non-perturbative effects.
The framework and data used in this paper are available as open-source
and published material in~\cite{groomRL,groomRL_lib,groomRL_data}.%
\footnote{The code is available at
  \url{https://github.com/JetsGame/GroomRL}, along with a C++ library
  at
  \url{https://github.com/JetsGame/libGroomRL}.}

\begin{algorithm}[tb]
  \caption{Grooming}
  \label{alg:grooming}
  \begin{algorithmic}
    \State {\bfseries Input:} policy $\pi_g$, binary tree node $\mathcal{T}^{(i)}$ 
    \State $a_t = \pi_g(\mathcal{T}^{(i)}\rightarrow s_t)$
    \If {$a_t==1$}
    \State $\mathcal{T}^{(j)} = \mathcal{T}^{(i)}$
    \While {$\mathcal{T}^{(j)} = (\mathcal{T}^{(j)}\rightarrow \text{parent})$}
    \State $\mathcal{T}^{(j)}\rightarrow s_t =(\mathcal{T}^{(j)}\rightarrow s_t)\,
    -\,(\mathcal{T}^{(i)}\rightarrow b\rightarrow s_t)$
    \EndWhile
    \State $\mathcal{T}^{(i)} = (\mathcal{T}^{(i)}\rightarrow a)$
    \State Grooming($\pi_g,\, \mathcal{T}^{(i)}$)
    \Else
    \State Grooming($\pi_g,\, \mathcal{T}^{(i)}\rightarrow a$)
    \State Grooming($\pi_g,\, \mathcal{T}^{(i)}\rightarrow b$)
    \EndIf
  \end{algorithmic}
\end{algorithm}

\section{Jet representation}
\label{sec:jet-rep}

Let us start by introducing the representation we use for jets.
We take the particle constituents of a jet, as defined by any modern
algorithm, and recombine them using a Cambridge/Aachen (CA) sequential
clustering algorithm~\cite{Dokshitzer:1997in,Wobisch:1998wt}.
The CA algorithm does a pairwise recombination, adding together the
momenta of the two particles with the closest distance as defined by
the measure
\begin{equation}
  \label{eq:CA-alg}
  \Delta^2_{ij} = (y_i - y_j)^2 + (\phi_i - \phi_j)^2\,,
\end{equation}
where $y_i$ is the rapidity, a measure of relativistic velocity along
the beam axis, and $\phi_i$ is the azimuthal angle of particle $i$
around the same axis.
This clustering sequence is then used to recast the jet as a full binary
tree, where each of the nodes contains information about the kinematic
properties of the two parent particles.
For each node $i$ of the tree we define an object $\mathcal{T}^{(i)}$
containing the current observable state $s_t$, as well as a pointer to
the two children nodes and one to the parent node.
The children nodes $a$ and $b$ are ordered in transverse momentum such
that $p_{t,a}>p_{t,b}$, and we label $a$ the ``harder'' child and $b$
the ``softer'' one.
The set of possible states is defined by a five-dimensional box, such
that the state of the node is a tuple
\begin{equation}
  \label{eq:node-tuple}  
  s_t=\left\{z, \Delta_{ab}, \psi, m, k_t\right\}\,,
\end{equation}
where $z=p_{t,b}/(p_{t,a}+p_{t,b})$ is the momentum fraction of the
softer child $b$,
$\psi=\tan^{-1}\big(\tfrac{y_b-y_a}{\phi_a-\phi_b}\big)$ is the
azimuthal angle around the $i$ axis, $m$ is the mass, and
$k_t= p_{t,b}\Delta_{ab}$ is the transverse momentum of $b$ relative
to $a$.

\subsection{Grooming algorithm}
\label{sec:groom-alg}

A grooming algorithm acting on a jet tree can be defined by a simple
recursive procedure which follows each of the branches and uses a
policy $\pi_g(s_t)$ to decide based on the values of
the current tuple $s_t$ whether to remove the softer of
the two branches.
This is shown in Algorithm~\ref{alg:grooming}, where the minus sign is
understood to mean the update of the kinematics of a node after
removal of a soft branch.
The grooming policy $\pi_g(s_t)$ returns an action $a_t\in\{0,1\}$,
with $a_t=1$ corresponding to the removal of a branch, and $a_t=0$
leaving the node unchanged.
The state $s_t$ is used to evaluate the current action-values
$Q^*(s,a)$ for each possible action, which in turn are used to
determine the best action at this step through a greedy policy.

\begin{figure}
  \centering
  \includegraphics[width=1.0\linewidth]{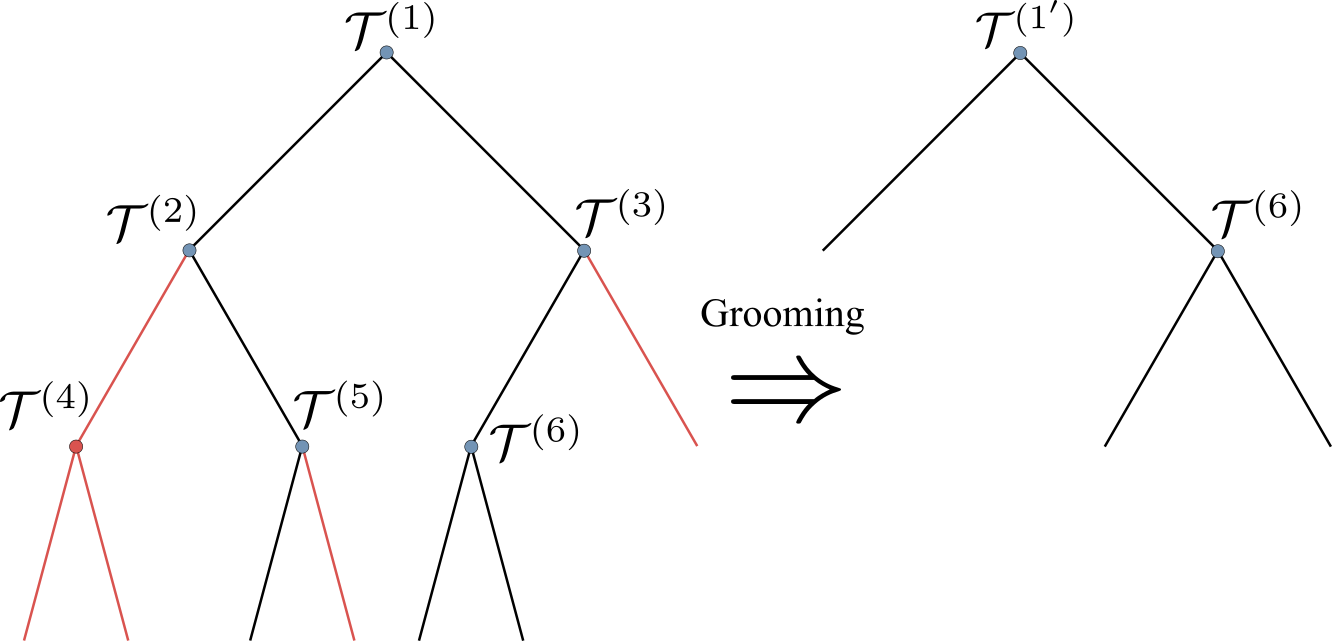}
  \caption{Example of grooming on the binary tree representation of a
    jet with the resulting tree after applying
    Algorithm~\ref{alg:grooming} shown on the right.
    Groomed branches are indicated in red, and the corresponding nodes
    have been removed on the right-hand side.}
  \label{fig:tree}
\end{figure}

An example of the action of a grooming algorithm on a tree is shown in
figure~\ref{fig:tree}, where the groomed branches are indicated in red.
The tree nodes whose kinematics have been modified by the removal of a
branch are indicated with a prime.

It is easy to translate modern grooming algorithms in this language.
For example, Recursive Soft Drop (RSD)~\cite{Dreyer:2018tjj}
corresponds to a policy
\begin{equation}
  \label{eq:RSD-alg}
  \pi_\text{RSD}(s_t) =
  \begin{cases}
    0\quad\text{if} \quad z > z_\text{cut} \big(\frac{\Delta_{ab}}{R_0}\big)^\beta\,,\\
    1\quad \text{else}\,,
  \end{cases}
\end{equation}
where $z_\text{cut}$, $\beta$ and $R_0$ are the parameters of the
algorithm, and $1$ corresponds as before to the action of removing the
tree branch with smaller transverse momentum.

\section{Setting up a grooming environment}
\label{sec:groomenv}

In order to find an optimal grooming policy $\pi_g$, we introduce an
environment and a reward function, formulating the problem in a way
that can be solved using a RL algorithm.

We initialize a list of all trees used for the training, from which a
tree is randomly selected at the beginning of each episode.
We then start by adding the root of the current tree to an empty priority queue,
which orders the nodes it contains according to their $\Delta_{ab}$
value.%
\footnote{This is not strictly necessary for a fully recursive
  algorithm, but allows for easier extensions to fixed depth
  algorithms such as the modified Mass Drop
  Tagger~\cite{Dasgupta:2013ihk} and Soft Drop~\cite{Larkoski:2014wba}.}

Each step consists in removing the first node from the priority queue,
and taking an action on which of its branches to keep based on the
state $s_t$ of that node.
Once a decision has been taken on the removal of the softer
branch, and the parent nodes have been updated accordingly, the
remaining children of the node are added to the priority
queue.
The reward function is then evaluated using the current state of the
tree.
The episode terminates once the priority queue is empty.

The framework described here deviates from usual RL implementations in
that the range of possible states for any episode are fixed at the start.
The transition probability between states
$\mathcal{P}(s_{t+1}|s_t,a_t)$ therefore does not necessarily depend
very strongly on the action, although a grooming action can result in
the removal of some of the future states and will therefore still have
an effect on the distribution.

For our implementation, we have relied on the
\texttt{gym} v0.12.1~\cite{1606.01540} and
\texttt{keras-rl} v0.4.2~\cite{plappert2016kerasrl} libraries for the
reinforcement learning component, while the neural network is set up
using \texttt{keras} v2.2.4~\cite{chollet2015keras} with \texttt{TensorFlow}
v1.13.1~\cite{tensorflow2015-whitepaper} as the backend.

\begin{figure}
  \centering
  \includegraphics[width=1.0\linewidth]{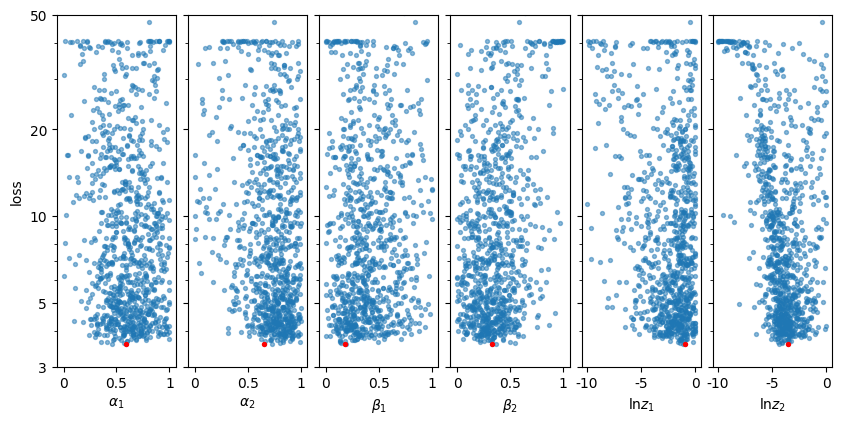}\\
  \includegraphics[width=0.5\linewidth,page=1]{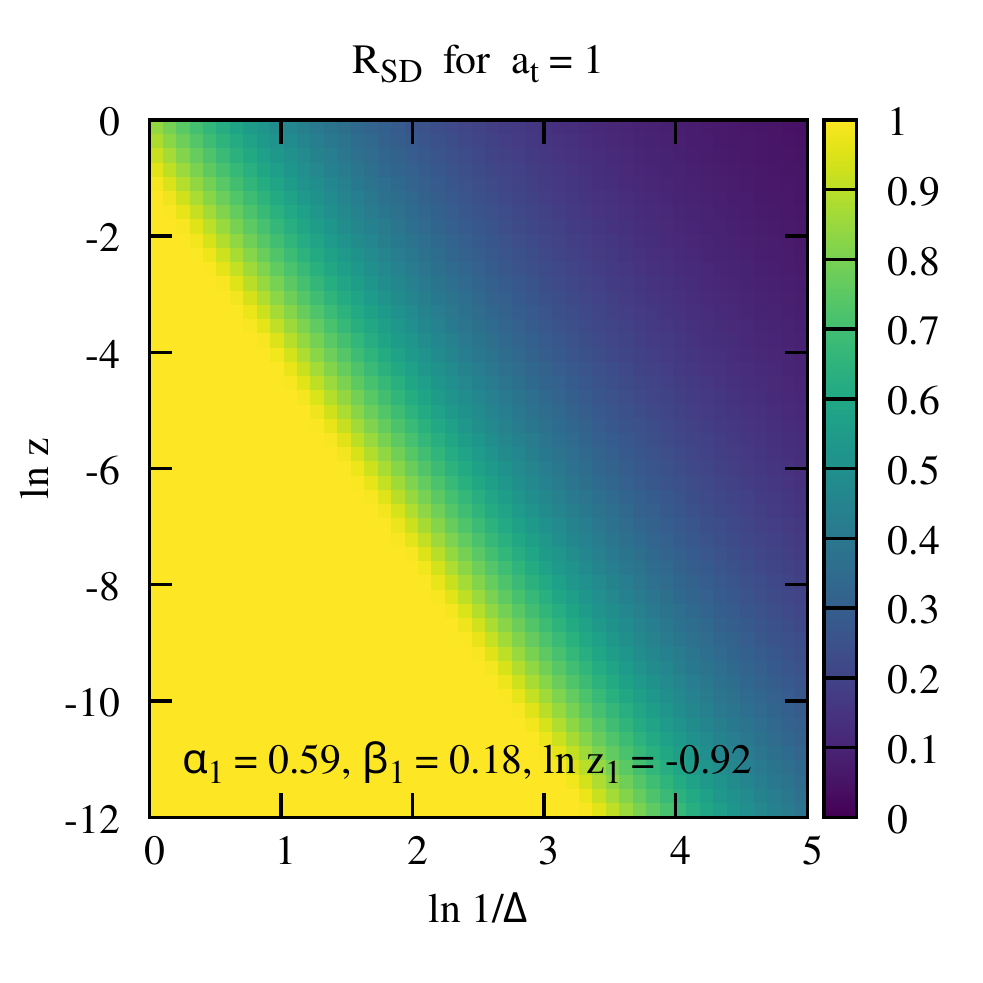}%
  \includegraphics[width=0.5\linewidth,page=2]{figures/SD-reward.pdf}%
  \caption{Top: Loss as a function of the reward parameters, with the
    optimal parameters shown in red.
    Bottom: Value of the two terms in the Soft-Drop reward function
    given in equation~(\ref{eq:SD-reward}) as a function of $\Delta$
    and $z$.}
  \label{fig:SD-reward}
\end{figure}

\subsection{Finding optimal hyper-parameters}
\label{sec:hyperopt}

The optimal choice of hyper-parameters, both for the model
architecture and for the grooming parameters, is determined using the
distributed asynchronous hyper-parameter optimization library
\texttt{hyperopt}~\cite{Bergstra:2013:MSM:3042817.3042832}.

The performance of an agent is evaluated by defining a loss function,
which is evaluated on a distinct validation set consisting of 50~000
signal and background jets.
For each sample, we evaluate the jet mass after grooming of each jet
and derive the corresponding distribution.
To calculate the loss function $\mathcal{L}$, we start by determining
a window $(w_\text{min},w_\text{max})$ containing a fraction $f=0.6$
of the final jet masses of the groomed signal distribution, defining
$w_\text{med}$ as the median value on that interval.
The loss function is then defined as
\begin{equation}
  \label{eq:loss-func}
  \mathcal{L} = \frac15|w_\text{max} - w_\text{min}|
  + |m_\text{target}-w_\text{med}|
  + 20 f_\text{bkg}\,,
\end{equation}
where $f_\text{bkg}$ is the fraction of the groomed background sample
contained in the same interval, and $m_\text{target}$ is a reference
value for the signal.

We scan hyper-parameters using 1000 iterations and select the
ones for which the loss $\mathcal{L}$ evaluated on the validation set
is minimal.
In practice we will do three different scans: to determine the best
parameters of the reward function, to find an optimal grooming
environment, and to determine the architecture of the DQN agent.
The scan is performed by requiring \texttt{hyperopt} to use a uniform
search space for continuous parameters, a log-uniform search space for
the learning rate and a binary choice for all integer or boolean
parameters.
The optimization used in all the results presented in this work
rely on the Tree-structured Parzen Estimator (TPE) algorithm.

\begin{figure*}
  \centering
  \includegraphics[width=1.0\linewidth]{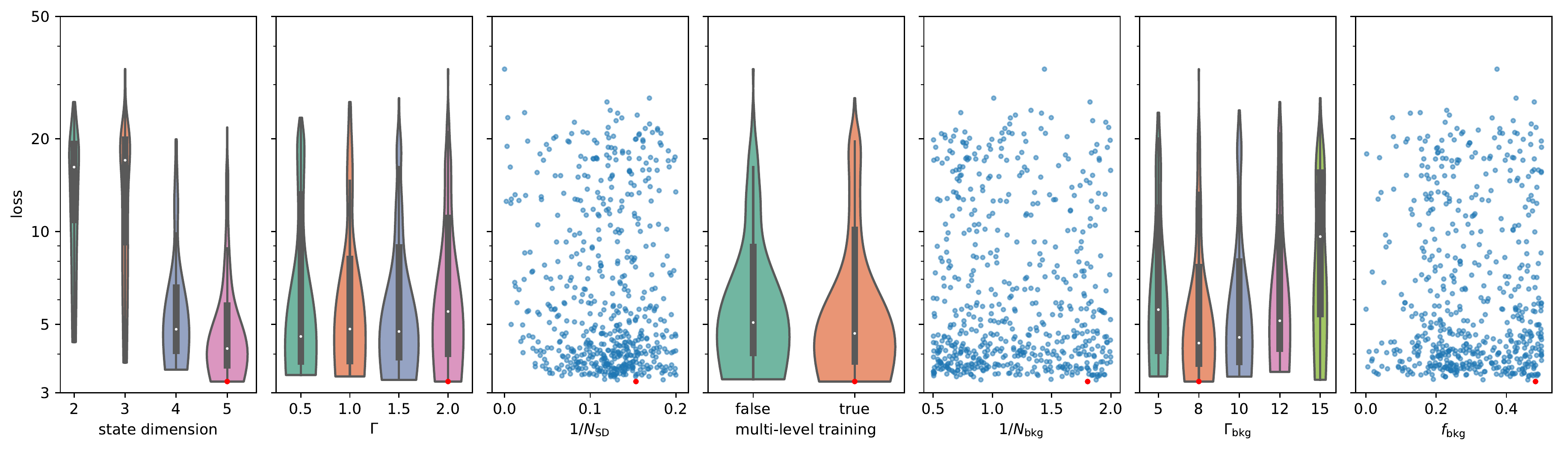}
  \caption{Distribution of the loss value for different grooming parameters.
    The best performing model is indicated in red.}
  \label{fig:groom-scan}
\end{figure*}

\subsection{Defining a reward function}
\label{sec:def-reward}

One of the key ingredients for the optimization of the grooming policy
is the reward function used at each step during the training.
We consider a reward with two components: a first piece evaluated
on the full tree, and another that considers only the kinematics of
the current node.

The first component of the reward compares the mass of the current jet
to a set target mass, typically the mass of the underlying boosted
object.
We implement this mass reward using a Cauchy distribution, which has
two free parameters, the target mass $m_\text{target}$ and a width
$\Gamma$, so that
\begin{equation}
  \label{eq:mass-reward}
  R_M(m) = \frac{\Gamma^2}{\pi(|m - m_\text{target}|^2 + \Gamma^2)}\,.
\end{equation}
Separately, we calculate a reward on the current node which gives a
positive reward for the removal of wide-angle soft radiation, as well
as for leaving intact hard-collinear emissions. This provides a 
baseline behavior for the groomer.
We label this reward component ``Soft-Drop'' due to its similarity
with the Soft Drop condition~\cite{Larkoski:2014wba}, and implement it
through exponential distributions
\begin{multline}
  \label{eq:SD-reward}
  R_\text{SD}(a_t, \Delta, z) =
  a_t\min\big(1,e^{-\alpha_1 \ln(1/\Delta) + \beta_1\ln (z_1/z)}\big)\\
  +(1-a_t)\max\big(0,1 - e^{-\alpha_2 \ln(1/\Delta) + \beta_2 \ln (z_2/z)}\big)\,,
\end{multline}
where $a_t=0,1$ is the action taken by the policy, and
$\alpha_{i}, \beta_{i}, z_{i}$ are free parameters.
The two terms determining $R_\text{SD}$ are shown in the lower panel
of figure~\ref{fig:SD-reward}, using parameter values determined
through asynchronous hyper-parameter optimization, shown in the upper
row of the figure.

The total reward function is then given by
\begin{equation}
  \label{eq:total-reward}
  R(m, a_t, \Delta, z) = R_M(m) +
    \frac{1}{N_\text{SD}} R_\text{SD}(a_t, \Delta, z)\,.
\end{equation}
Here $N_\text{SD}$ is a normalization factor determining the weight
given to the second component of the reward.

\subsection{RL implementation and multi-level training}
\label{sec:multi-level}

For the applications in this article, we have implemented a DQN agent
that contains a groomer module, which is defined by the underlying NN
model and the test policy used by the agent.
The groomer can be extracted after the model has been trained, using a
greedy policy to select the best action based on the $Q$-values
predicted by the NN.
This allows for straightforward application of the resulting grooming
strategy on new samples.

The training sample consists of 500~000 signal and background jets
simulated using \texttt{Pythia} 8.223~\cite{Sjostrand:2014zea}.
We will construct two separate models by considering two signal
samples, one with boosted $W$ jets and one with boosted top jets,
while the background always consists of QCD jets.
We use the $WW$ and $t\bar{t}$ processes, with hadronically decaying
$W$ and top, to create the signal samples, and the dijet process for
the background.
Jets are clustered using the anti-$k_t$
algorithm~\cite{Cacciari:2008gp,Cacciari:2011ma} with radius $R=1.0$,
and are required to pass a selection cut, with transverse momentum
$p_t > 500$ GeV and rapidity $|y|<2.5$.
All samples used in this article are available
online~\cite{groomRL_data}.
The grooming environment is initialized by reading in the training
data and creating an event array containing the corresponding jet
trees.

To train the RL agent, we use a multi-level approach taking into
account both signal and background samples.
At the beginning of each episode, we select either a signal jet or a
background jet, with probability $1-p_\text{bkg}$.
For signal jets, the reward function uses a reference mass set to the
$W$-boson mass, $m_\text{target}=m_W$, or to the top mass,
$m_\text{target} = m_t$, depending on the choice of sample.
In the case of the background the mass reward function in
equation~(\ref{eq:total-reward}) is changed to
\begin{equation}
  \label{eq:mass-reward-bkg}
  R^{\rm bkg}_M(m) = \frac{m}{\Gamma_{\rm bkg}}
  \exp\Big(-\frac{m}{\Gamma_{\rm bkg}}\,\Big)\,.
\end{equation}
The width parameters $\Gamma$, $\Gamma_{\rm bkg}$ are also set to different values for signal and
background reward functions, and are determined through a
hyper-parameter scan.

We found that while this multi-level training only marginally improves
the performance, it noticeably reduces the variability of the model.

\subsection{Determining the RL agent}
\label{sec:determine-agent}

The DQN agent uses an Adam
optimizer~\cite{DBLP:journals/corr/KingmaB14}, and the training is
performed with a Boltzmann policy, which chooses an action according
to weighted probabilities, with the current best action being the
likeliest.

Let us now determine the remaining parameters of the DQN agent.
To this end, we perform two independent scans, for the
grooming environment and for the network architecture.

\begin{figure*}
  \centering
  \includegraphics[width=1.0\linewidth]{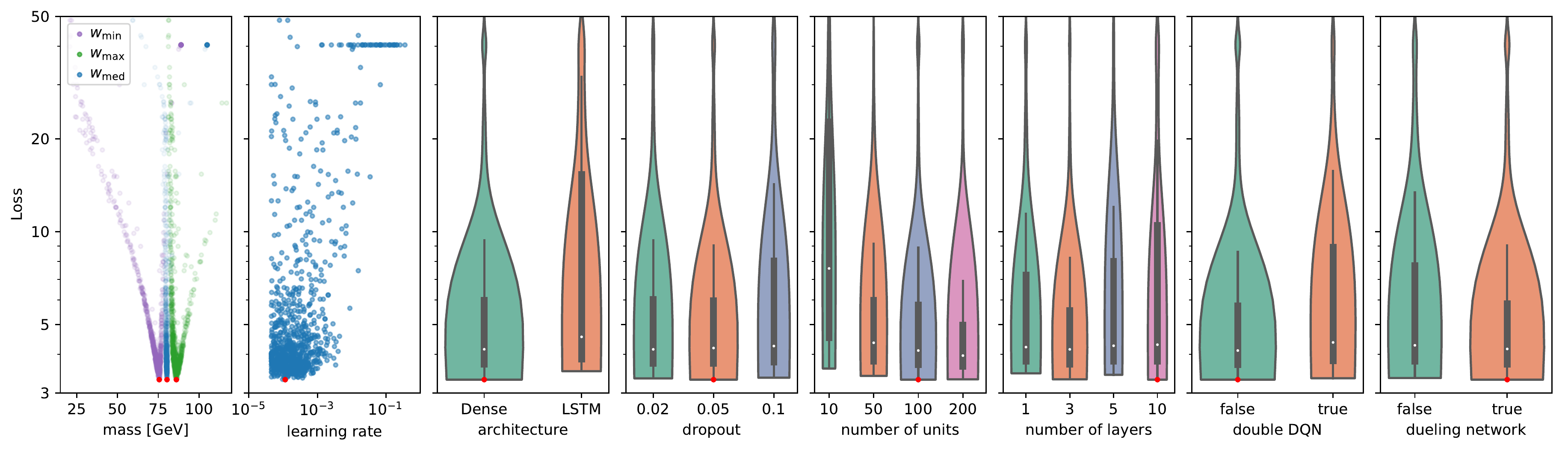}
  \caption{Distribution of the loss value for different architecture configurations.
    The best performing model is indicated in red.}
  \label{fig:architecture-scan}
\end{figure*}

The grooming environment has several options, which are shown in
figure~\ref{fig:groom-scan}.
Here the distribution of loss values for discrete options are
displayed using violin plots, showing both the probability density
of the loss values as well as its quartiles.
The first plot is the dimensionality of the state observed at each
step, which can be a subset of the tuple given in
equation~(\ref{eq:node-tuple}).
We can observe that as the dimension of the input state is increased,
the NN is able to leverage this additional information,
leading to a decrease of the loss function.
The scan over the normalization parameters of the reward functions
shows that it is preferable to use a small width $\Gamma$ for the
signal, with a large value $\Gamma_{\rm bkg}$ for the background, as
well as a small value for the $1/N_{\rm SD}$ factor.
One can also see that the multi-level training described in
section~\ref{sec:multi-level} leads to a distribution of loss values
concentrated at smaller values.
We have also allowed for several functional forms of the signal mass
reward function, although for our final model we will use a Cauchy
distribution.

The parameters of the network architecture are shown in
figure~\ref{fig:architecture-scan}, with the first plot showing the
mass window containing $60\%$ of the signal distribution, with the
median of that interval shown in blue.
The scatter plot of the learning rate used for the Adam optimizer
shows that a value slightly above $10^{-4}$ yields the best result.
The scan shows a preference for a dense network with a large number of
units and layers as well as a dropout layer as the architecture of the
NN.
Finally, we see that using duelling
networks~\cite{DBLP:journals/corr/WangFL15} leads to a small
improvement of the model, while double
Q-learning~\cite{DBLP:journals/corr/HasseltGS15} does not.

\subsection{Optimal GroomRL model}
\label{sec:optimal-groomRL}

The final \texttt{GroomRL} model is trained using the full training
sample with 500~000 signal/background jets for 1~000~000 epochs. The overall
training time requires four hours of training using a single NVIDIA
GTX 1080 Ti GPU with 12 GB of memory which includes all the training
jet trees and the DQN parameters.

The parameters of the best \texttt{GroomRL} model obtained following
the strategy presented in the previous sections is listed in
table~\ref{tab:parameters}.
Here two values are given for the $m_{\rm target}$ parameter, which
are used to train on either a sample consisting of $W$ bosons or of
top quarks.
The resulting models are labeled \texttt{GroomRL-W} and
\texttt{GroomRL-Top} respectively.

\begin{table}
  \begin{center}
    \phantom{x}\medskip
    \begin{tabular}{ccc}
      \toprule
      \textbf{Parameters} && \textbf{Value}\\
      \midrule
      $m_{\rm target}$ && 80.385 GeV or 173.2 GeV\\
      \midrule
      $s_t$ dimension && 5\\ [3pt]
      reward && Cauchy\\ [3pt]
      $\Gamma$ && 2 GeV\\ [3pt]
      $(\alpha_{1},\beta_1,\ln z_1)$ && $(0.59,0.18,-0.92)$\\ [3pt]
      $(\alpha_{2},\beta_2,\ln z_2)$ && $(0.65,0.33,-3.53)$\\ [3pt]
      $1/N_{\rm SD}$ && 0.15\\ [3pt]
      multi-level training && Yes\\ [3pt]
      $\Gamma_{\rm bkg}$ && 8 GeV\\ [3pt]
      $1/N_{\rm bkg}$ && 1.8 or 1.0 \\ [3pt]
      $p_{\rm bkg}$ &&  0.48 or 0.2\\
      \midrule
      learning rate && $10^{-4}$\\ [3pt]
      Dueling NN && Yes\\ [3pt]
      Double DQN && No\\ [3pt]
      Policy && Boltzmann\\ [3pt]
      $N_{\rm epochs}^{\rm max}$ && 500~000\\ [3pt]
      Architecture && Dense\\ [3pt]
      Dropout && 0.05\\ [3pt]
      Layers && 10\\ [3pt]
      Nodes && 100\\ [3pt]
      Optimizer && Adam \\
      \bottomrule
    \end{tabular}
    \caption{Final parameters for \texttt{GroomRL}, with the two
      values of $m_{\rm target}$ corresponding to the $W$ and top
      mass.}
    \label{tab:parameters}
  \end{center}
\end{table}

\begin{figure}
  \centering
  \includegraphics[scale=0.5]{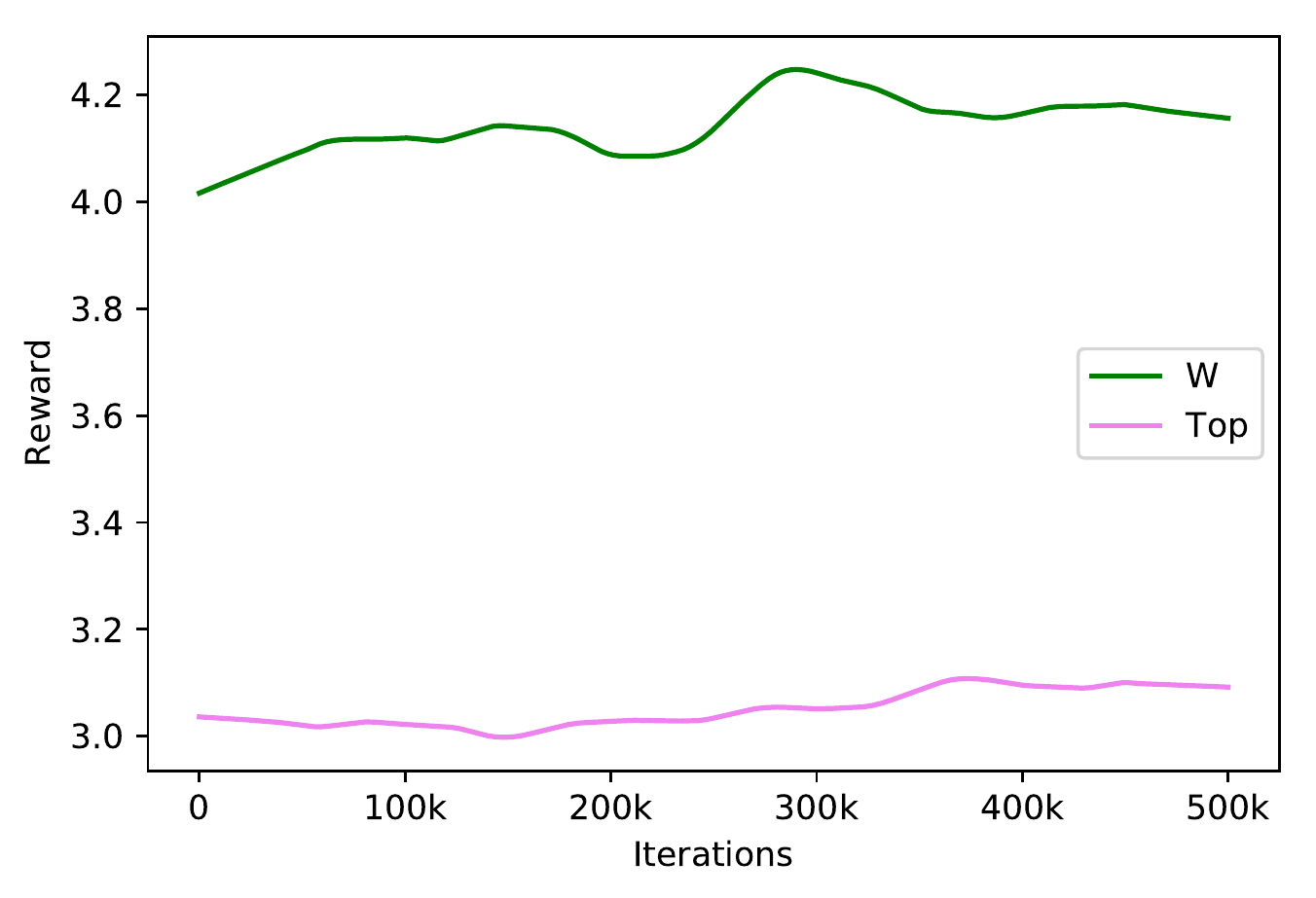}
  \caption{Reward evolution during training of the \texttt{GroomRL}
    on $W$ and top data.
    A LOESS smoothing is applied to the original curves.}
  \label{fig:reward}
\end{figure}

In figure~\ref{fig:reward} we show the reward value during the
training of the \texttt{GroomRL} for $W$ bosons and top quarks, after
applying the LOESS smoothing algorithm on the original curve. We
observe an improvement of the reward function during the first 300~000
training epochs, with the reward becoming relatively stable after that
point.

\subsection{Alternative approaches}
\label{sec:alternative}
In this section, we have introduced a novel implementation of RL to
tackle the problem of tree pruning.
A number of alternative methods could be studied to approach this
problem, most notably Monte-Carlo Tree Search (MCTS)
algorithms~\cite{coulom2006mcts,kocsis2006uct} and binary classifiers.
The heuristic search methods from MCTS explore the tree through random
sampling, taking random actions to progress through the tree.
Once an endpoint is reached, the result is used to weight the nodes
and improve future decisions.

More recently, a NN based MCTSnet implementation was
proposed~\cite{Guez2018LearningTS}, which introduces a framework to
learn how to search the tree, integrating simulation-based planning
into a NN.

These techniques might provide an interesting basis to construct an
efficient groomer.
However due to the wide variability of the trees considered in our
case study, where each new episode starts from a unique tree, this
would require a substantial modification of the algorithm.

Alternatively, one could use a contextual bandit
solver~\cite{Auer:2003:NMB:589343.589365,DBLP:journals/corr/AgarwalHKLLS14}
to train a jet grooming policy.
We would expect this method to yield similar results, however, this
method does not allow for the modification of the future nodes by the
current grooming decision, and is not as easily extendable as our
current framework.

Finally, one could attempt to build a jet grooming algorithm
from a binary classifier, which uses an input state to determine which
action to take next.
The main drawback of this method is that one can not straightforwardly
impose as loss function the mass resolution of the tree, as this
depends on previous states of the current episode.
As such, the problem we consider is particularly well adapted to a
RL approach.

We leave a more thorough study of the application of these alternative
tools to jet grooming for future work.

\section{Jet mass spectrum}
\label{sec:jetmass}

Let us now apply the {\tt GroomRL} models defined in
section~\ref{sec:optimal-groomRL} to new data samples.
We consider three test sets of 50~000 elements each: one with QCD jets,
one with $W$ initiated jets and one with top jets.
The size of the window containing $60\%$ of the mass spectrum of the
$W$ sample, as well as the corresponding median value, are given in
table~\ref{tab:window} for each different grooming strategy.
As a benchmark, we compare to the RSD algorithm, using parameters
$z_\text{cut}=0.05$, $\beta=1$ and $R_0=1$.
One can notice a sizeable reduction of the window size after grooming
with the machine learning based algorithms, while all groomers are
able to reconstruct the peak location to a value very close to the $W$
mass.

\begin{figure*}
  \centering
  
  \subfloat[QCD]{\includegraphics[width=0.33\textwidth]{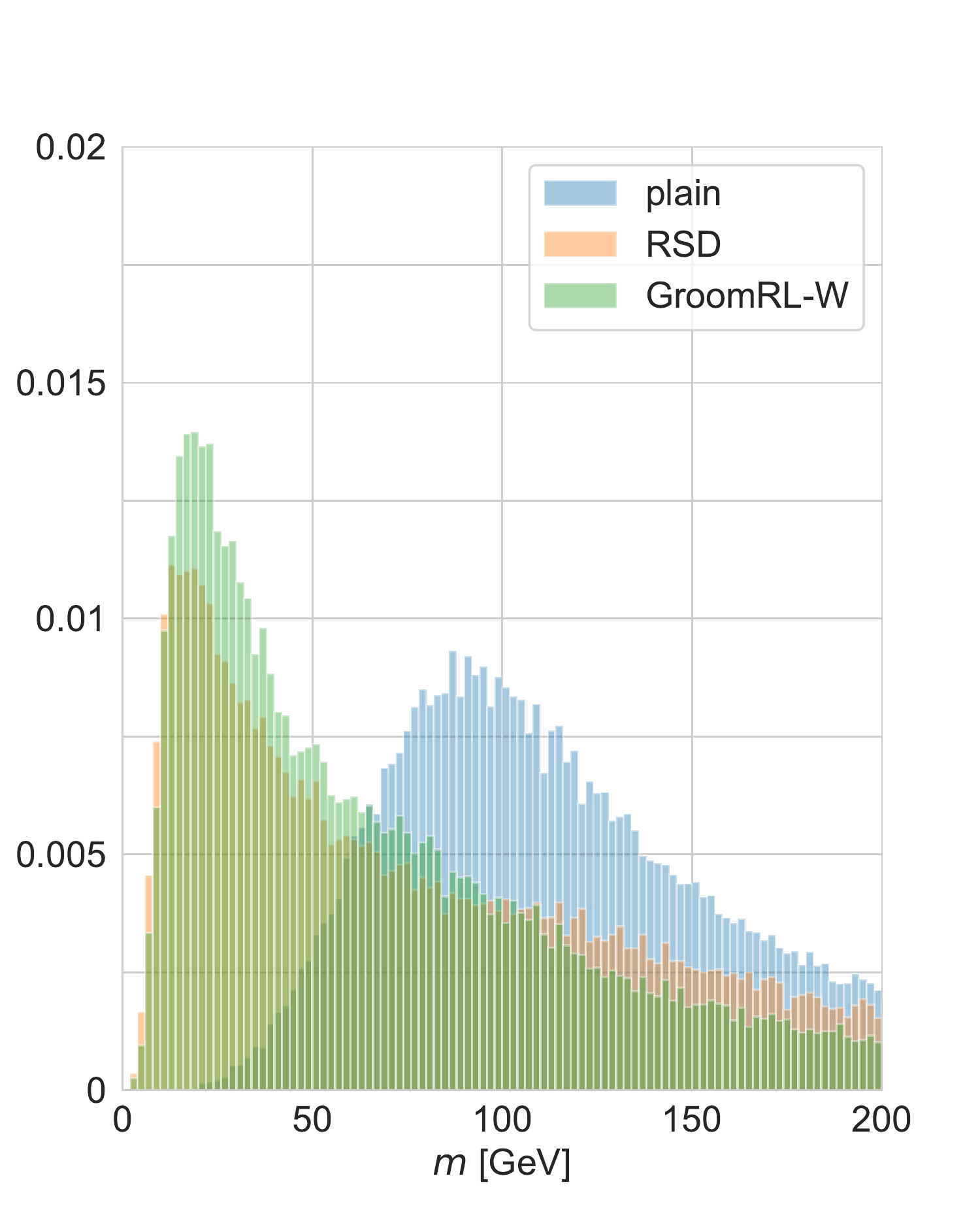}%
    \label{fig:qcd_jetmass}}%
  \subfloat[W]{\includegraphics[width=0.33\textwidth]{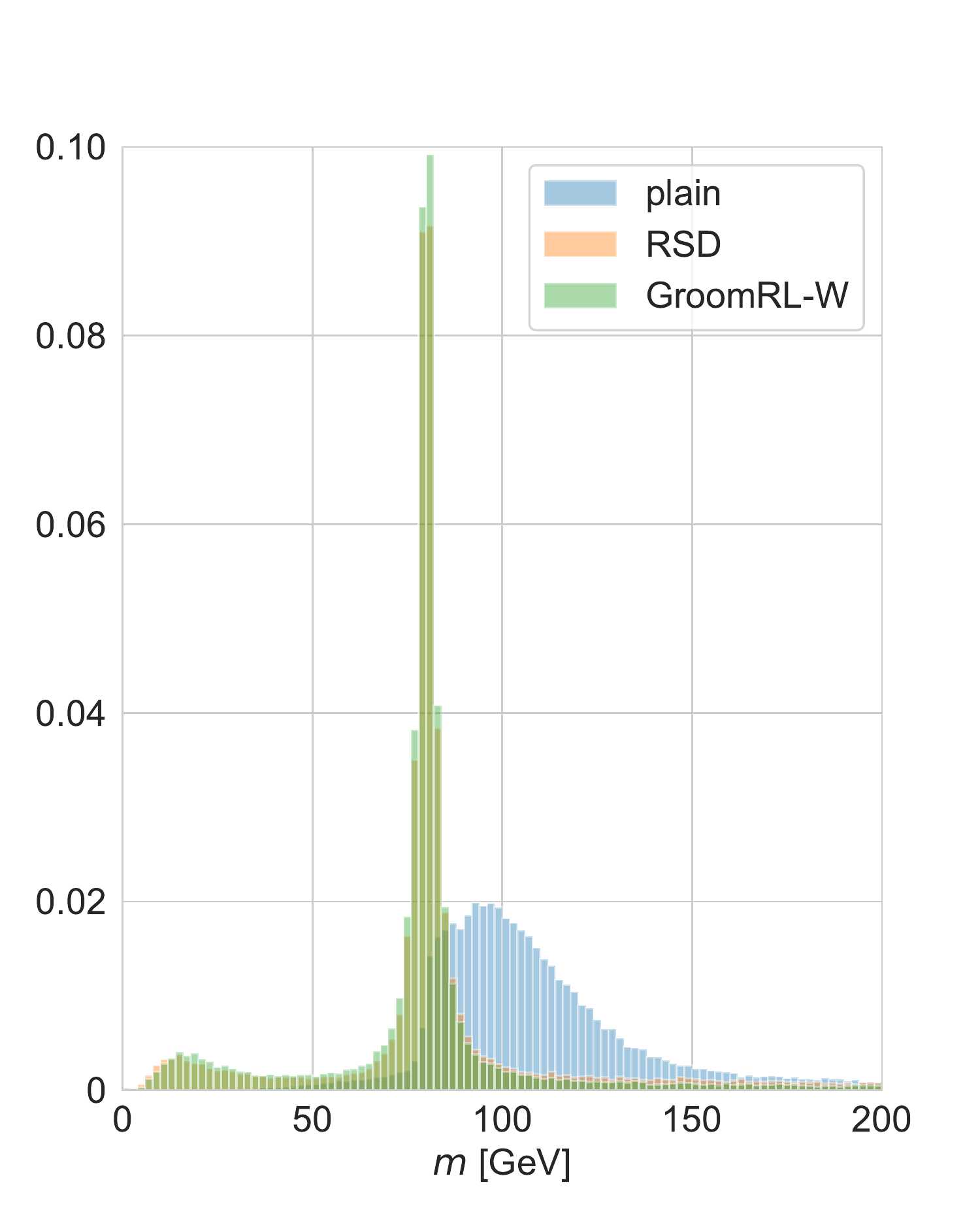}%
    \label{fig:W_jetmass}}%
  \subfloat[top]{\includegraphics[width=0.33\textwidth]{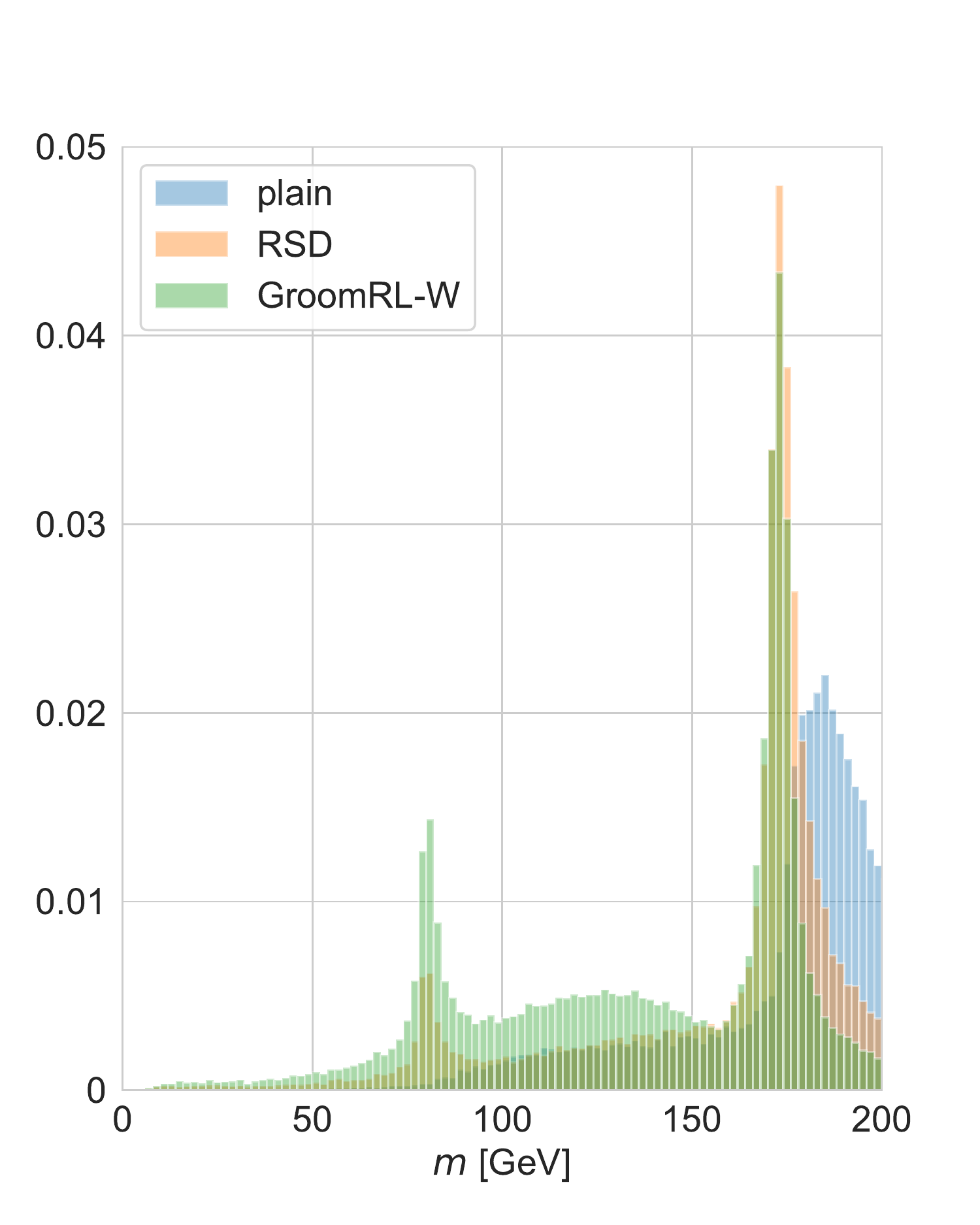}%
    \label{fig:top_jetmass}}%
  \caption{Groomed jet mass spectrum for (a) QCD jets, (b) $W$ jets,
    (c) top jets. The \texttt{GroomRL-W} curve is obtained from
    training on $W$ data.}
  \label{fig:jet-mass}
\end{figure*}

\begin{figure*}
  \centering
  
  \subfloat[QCD]{\includegraphics[width=0.33\textwidth]{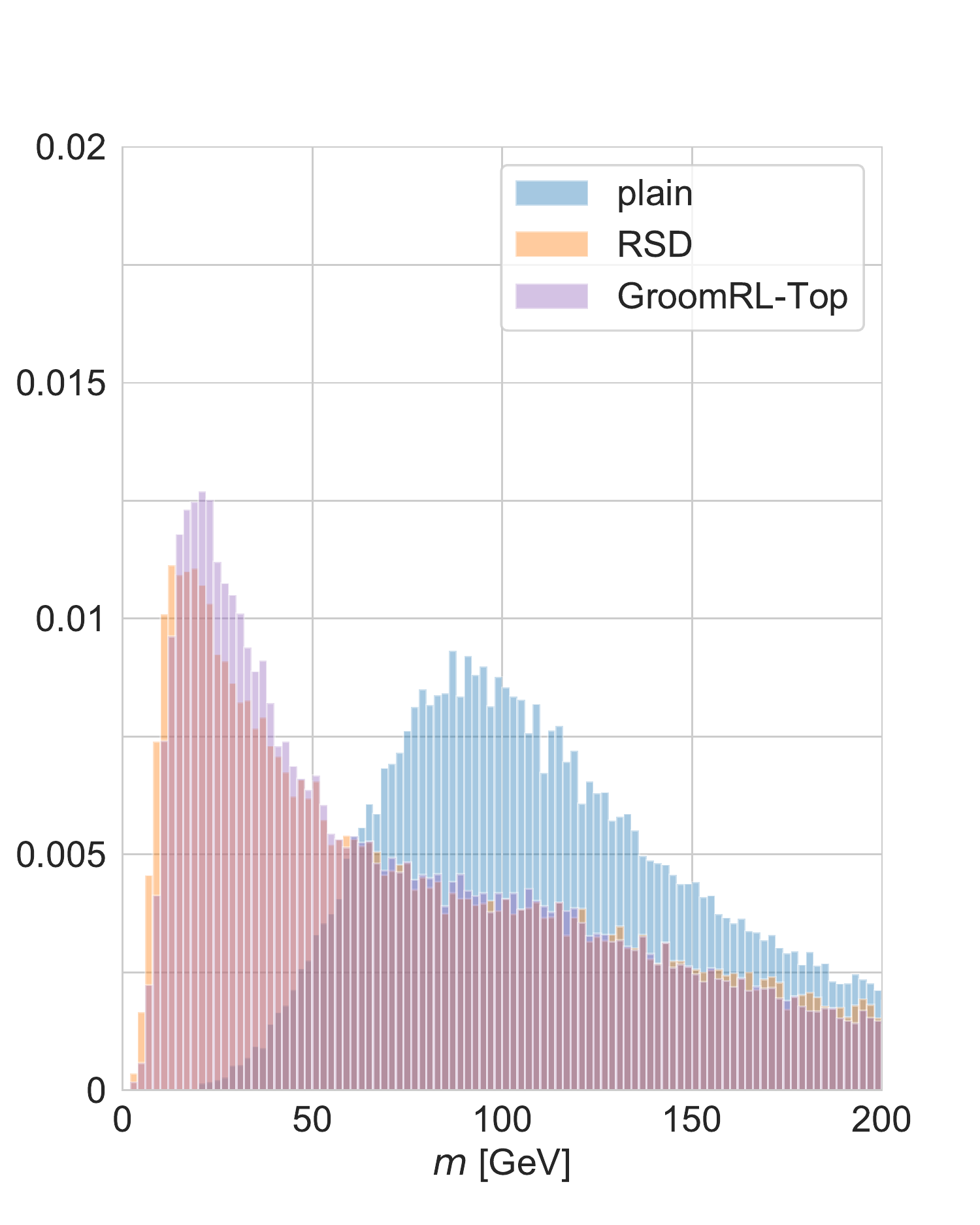}%
    \label{fig:qcd_jetmass_top}}%
  \subfloat[W]{\includegraphics[width=0.33\textwidth]{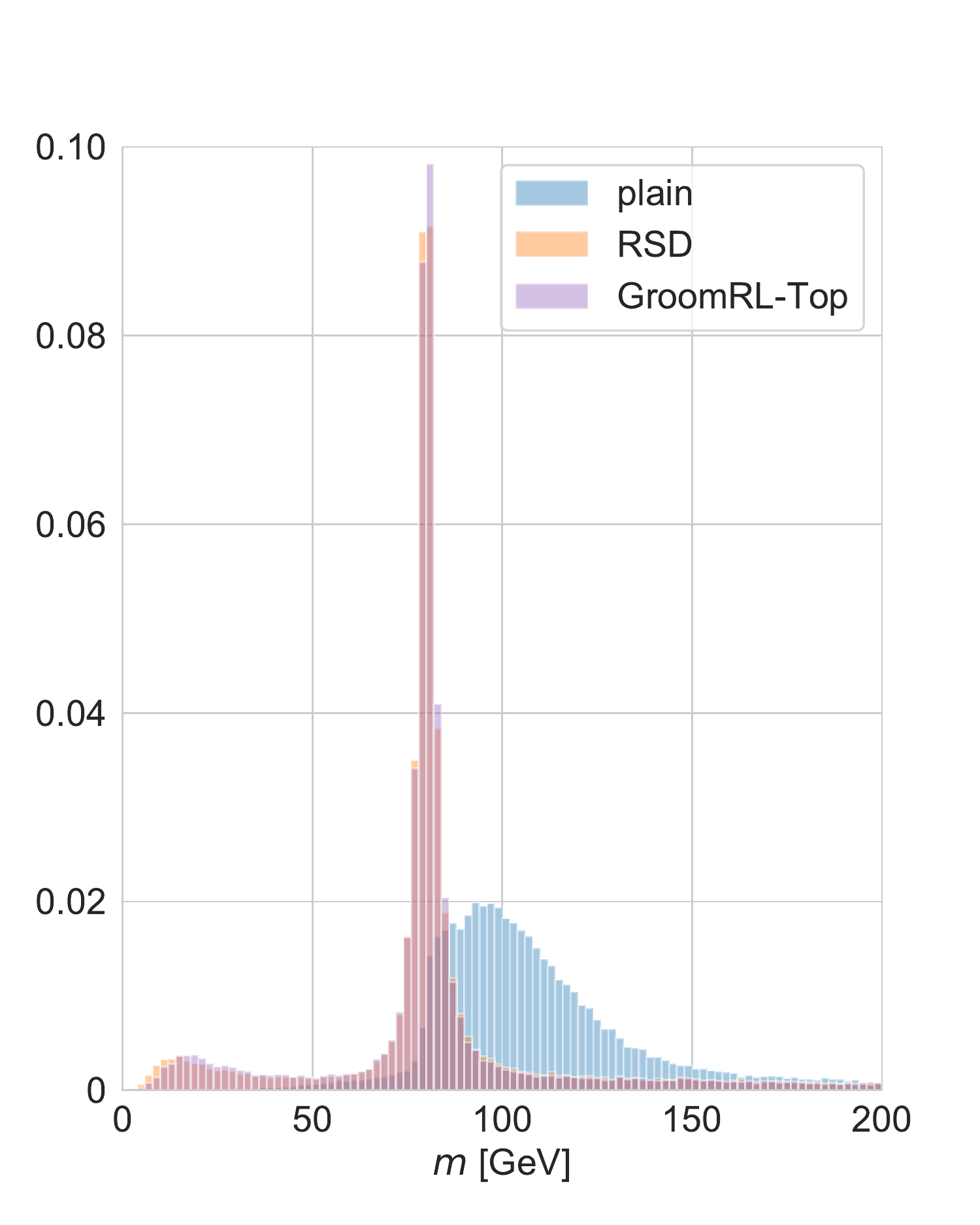}%
    \label{fig:W_jetmass_top}}%
  \subfloat[top]{\includegraphics[width=0.33\textwidth]{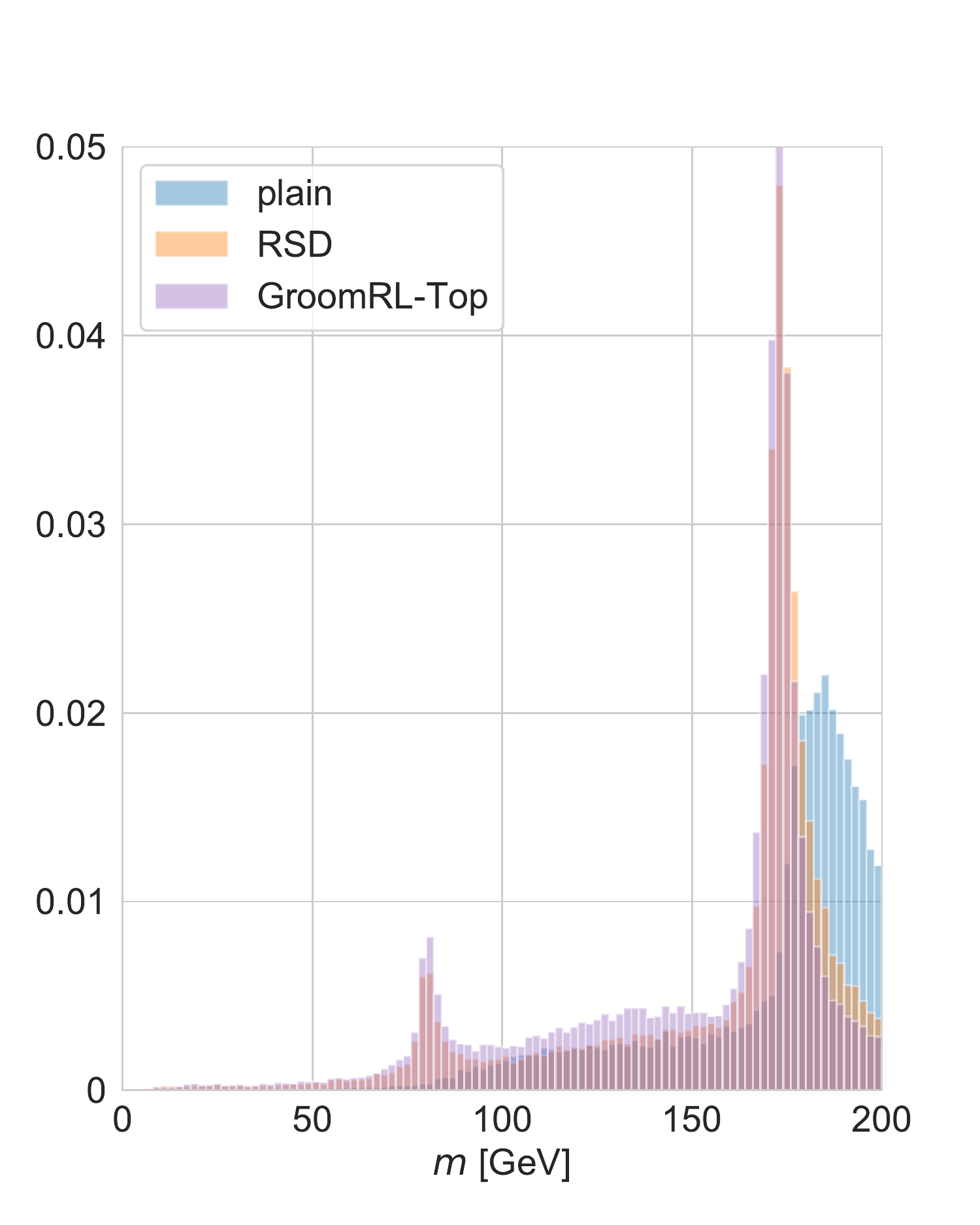}%
    \label{fig:top_jetmass_top}}%
  \caption{Groomed jet mass spectrum for (a) QCD jets, (b) $W$ jets,
    (c) top jets. The \texttt{GroomRL-Top} curve is obtained from
    training on top data.}
  \label{fig:jet-mass-top}
\end{figure*}

The distribution of the jet mass after grooming for each of these
samples is shown in figures~\ref{fig:jet-mass}
and~\ref{fig:jet-mass-top}.
Each curve gives the differential cross section $d\sigma/d m_j$
normalized by the total cross section.
Figure~\ref{fig:jet-mass} shows results for the grooming algorithm
trained on a $W$ sample, while the results of the algorithm trained on
top data are given in figure~\ref{fig:jet-mass-top}.
As references, the ungroomed (or plain) jet mass and the jet mass
after RSD grooming are also given, in blue and orange respectively.
As expected, one can observe that for the ungroomed case the
resolution is very poor, with the QCD jets having large masses due to
wide-angle radiation, while the $W$ and top mass peaks are heavily
distorted.
In contrast, after applying RSD or {\tt GroomRL}, the jet mass is
reconstructed much more accurately.
One interesting feature of {\tt GroomRL} is that it is able to lower
the jet mass for quark and gluon jets, further reducing the background
contamination in windows close to a heavy particle mass.

For the $W$ case, shown in figures~\ref{fig:W_jetmass}
and~\ref{fig:W_jetmass_top}, there is a sharp peak around the $W$ mass
$m_W$, with the {\tt GroomRL} method providing slightly better
resolution.
It is also particularly noteworthy that both the \texttt{GroomRL-W}
and the \texttt{GroomRL-Top} algorithms have similar performance,
despite the latter one having been trained on a completely different
data set. This demonstrates that the tools derived from our framework
are robust and can be applied to data sets beyond their training range
with good results.

\begin{table}
  \begin{center}
    \phantom{x}\medskip
    \begin{tabular}{lccccccc}
      \toprule
      && $w_\text{max} - w_\text{min}$ [GeV] && $w_\text{med}$ [GeV]\\
      \midrule
      plain && $44.65$ && $104.64$\\ [4pt]
      \texttt{GroomRL-W} && $10.70$ && $80.09$\\ [4pt]
      \texttt{GroomRL-Top} && $13.88$ && $80.46$ \\ [4pt]
      RSD && $16.96$ && $80.46$\\
      \bottomrule
    \end{tabular}
    \caption{Size of the window containing $60\%$ of the $W$ mass
      spectrum, and median value on that interval.}
    \label{tab:window}
  \end{center}
\end{table}

In top jets, displayed in figures~\ref{fig:top_jetmass}
and~\ref{fig:top_jetmass_top}, the enhancements are even more
noticeable.
Here again, the performance of both algorithms is similar, despite the
fact that the training of {\tt GroomRL-W} did not involve any top-related
data.

\begin{figure*}
  \centering
  \includegraphics[width=1.0\linewidth]{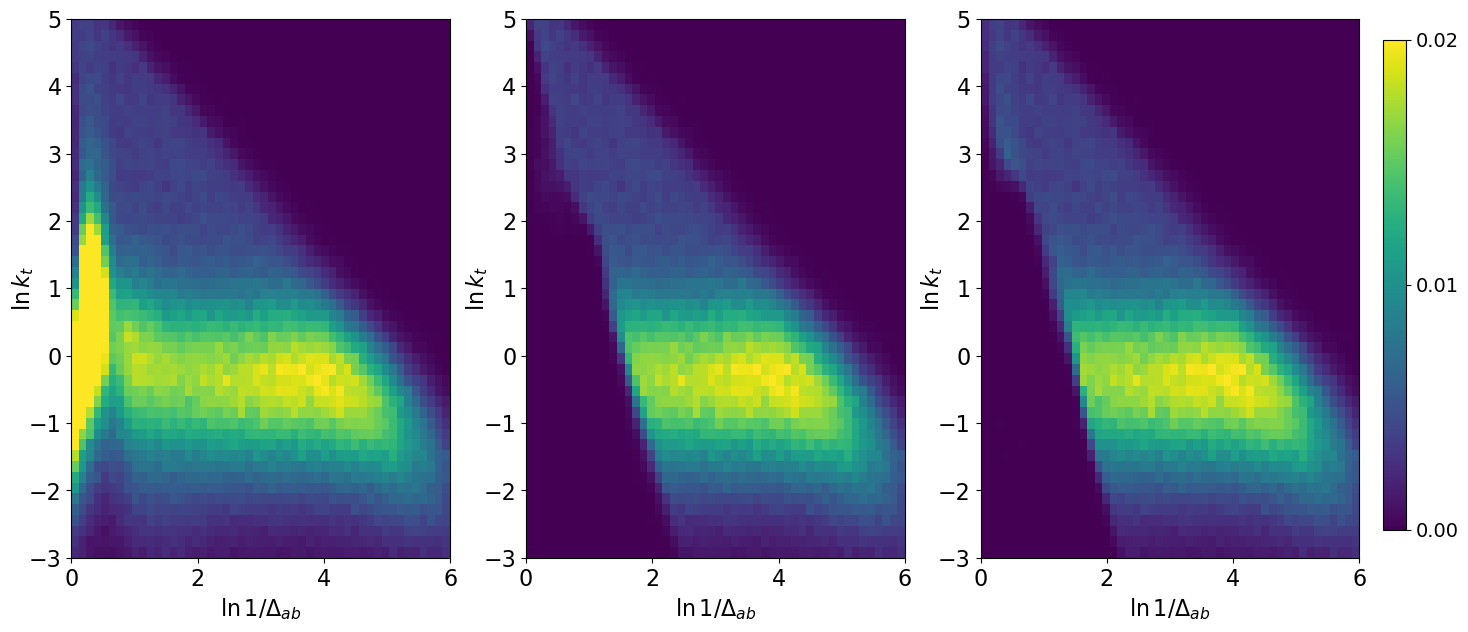}
  \begin{minipage}{0.31\linewidth}
    \begin{center}
      (a) Plain
    \end{center}
  \end{minipage}%
  \begin{minipage}{0.31\linewidth}
    \begin{center}
      (b) \texttt{GroomRL-W}
    \end{center}
  \end{minipage}%
  \begin{minipage}{0.31\linewidth}
    \begin{center}
      (c) \texttt{GroomRL-Top}
    \end{center}
  \end{minipage}\hfill
  \caption{Primary Lund jet plane density for QCD jets before (a)
    and after grooming with {\tt GroomRL} trained on $W$ (b) or
    top (c) samples.}
  \label{fig:lundplane}
\end{figure*}

Finally, in figure~\ref{fig:lundplane}, we show the primary Lund jet
plane density as defined in~\cite{Dreyer:2018nbf} after grooming 
with {\tt GroomRL-W} and {\tt GroomRL-Top}, averaged over 50~000 jets.
This gives a useful visualization of radiation patterns within a jet, 
providing a physical interpretation of the grooming behavior.
The primary Lund jet plane is defined through the
$(\ln 1/\Delta_{ab},\ln k_t)$ coordinates of each of the
states of the ``primary'' declustering sequence, i.e.\ traversing the
jet tree by successively following the hardest branch
$\mathcal{T}^{(i)}\rightarrow a$.
The upper boundary of the triangle is due to the kinematic limit of
emissions.
In contrast, the lower edge corresponds to radiation that gets removed
by the grooming algorithm, so that only sufficiently energetic or
collinear partons remain in the groomed jet.

An interesting feature of figure~\ref{fig:lundplane} is that can one
observe that despite producing similar jet mass spectra, the {\tt
  GroomRL-W} and {\tt GroomRL-Top} algorithms differ somewhat, with
the former retaining more radiation at wide angles than the latter.

\subsection{Robustness to non-perturbative effects}
\label{sec:np-eff}

Let us now consider the impact of non-perturbative effects such as
hadronization and underlying event on groomed jets.
A key feature of grooming algorithms such as mMDT and Soft Drop is
that they reduce the sensitivity of observables to non-perturbative
effects, allowing for precise comparisons between theoretical
predictions and experimental measurements.

To study the robustness of \texttt{GroomRL} to these contributions, we
consider three different QCD jet samples generated through
\texttt{Pythia}'s dijet process.
The first one, which we denote as ``truth-level'' and used already in
the previous sections, includes all non-perturbative effects.
A ``hadron-level'' sample is obtained by removing multiple parton
interactions from the simulation, and finally a ``parton-level''
sample is generated by further turning off the hadronization step in
\texttt{Pythia}.

\begin{figure*}
  \centering  
  \subfloat[Plain]{\includegraphics[width=0.45\textwidth]{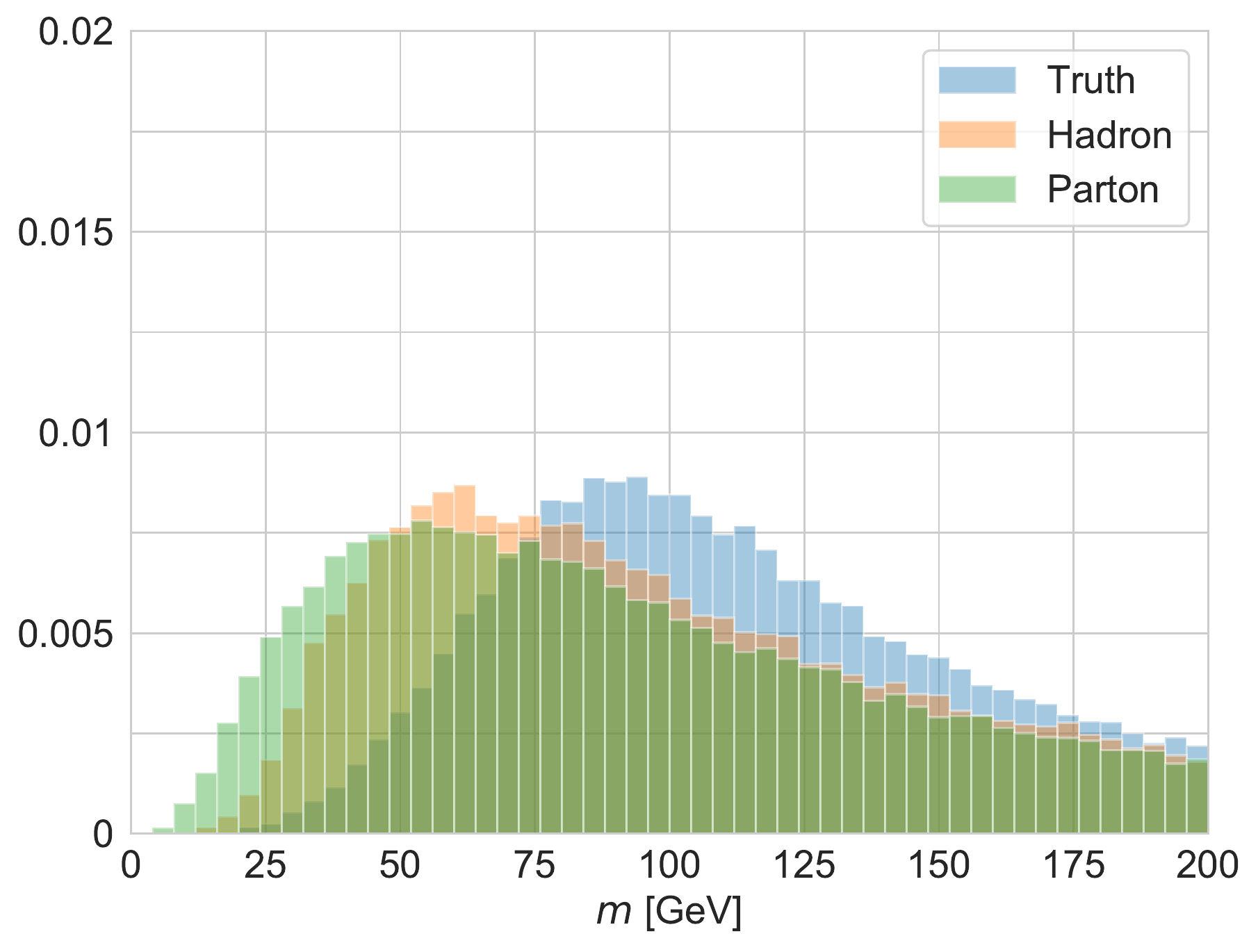}%
    \label{fig:np-effects-plain}}~%
  \subfloat[\texttt{GroomRL-W}]{\includegraphics[width=0.45\textwidth]{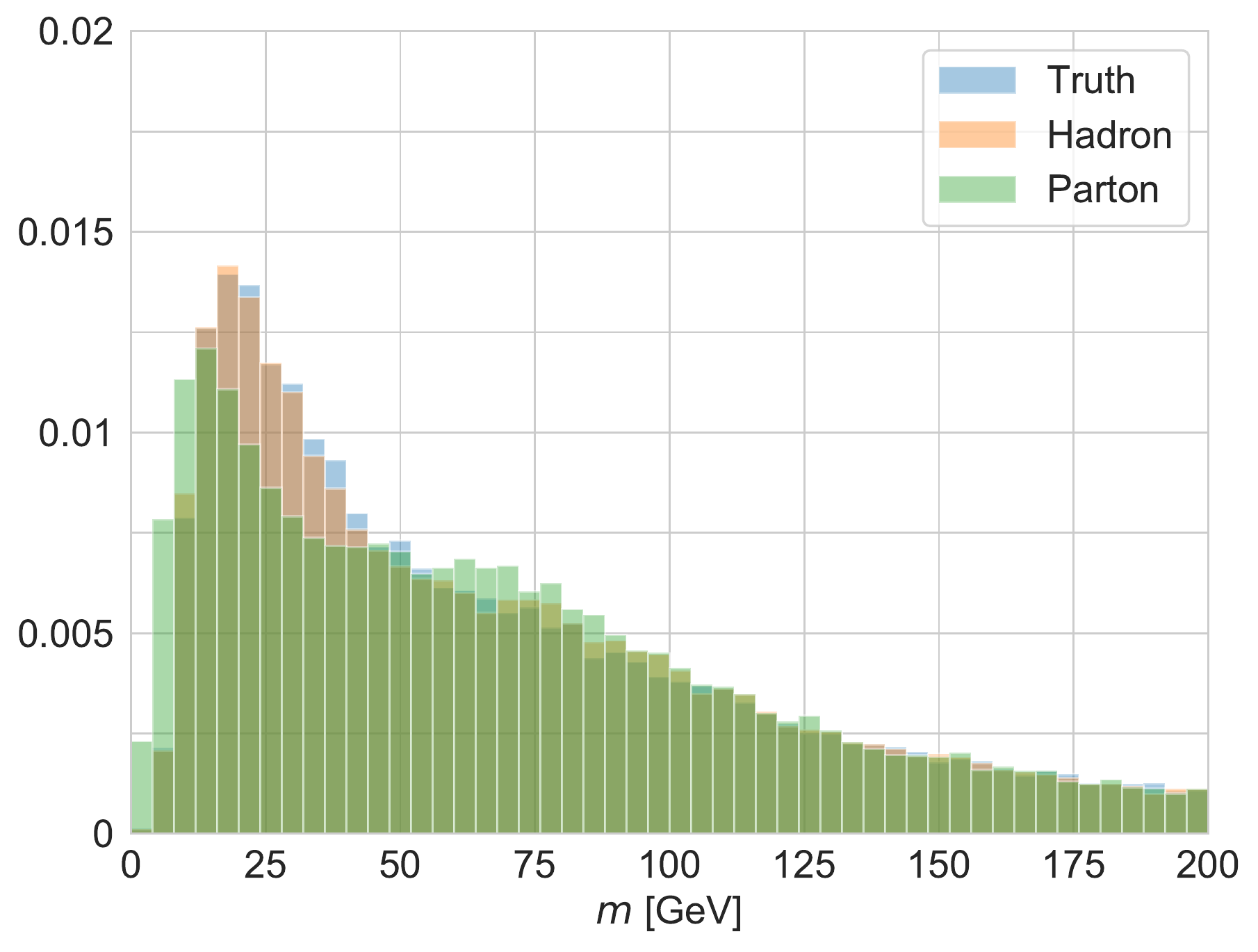}%
    \label{fig:np-effects-groomW}}\\
  \subfloat[\texttt{GroomRL-Top}]{\includegraphics[width=0.45\textwidth]{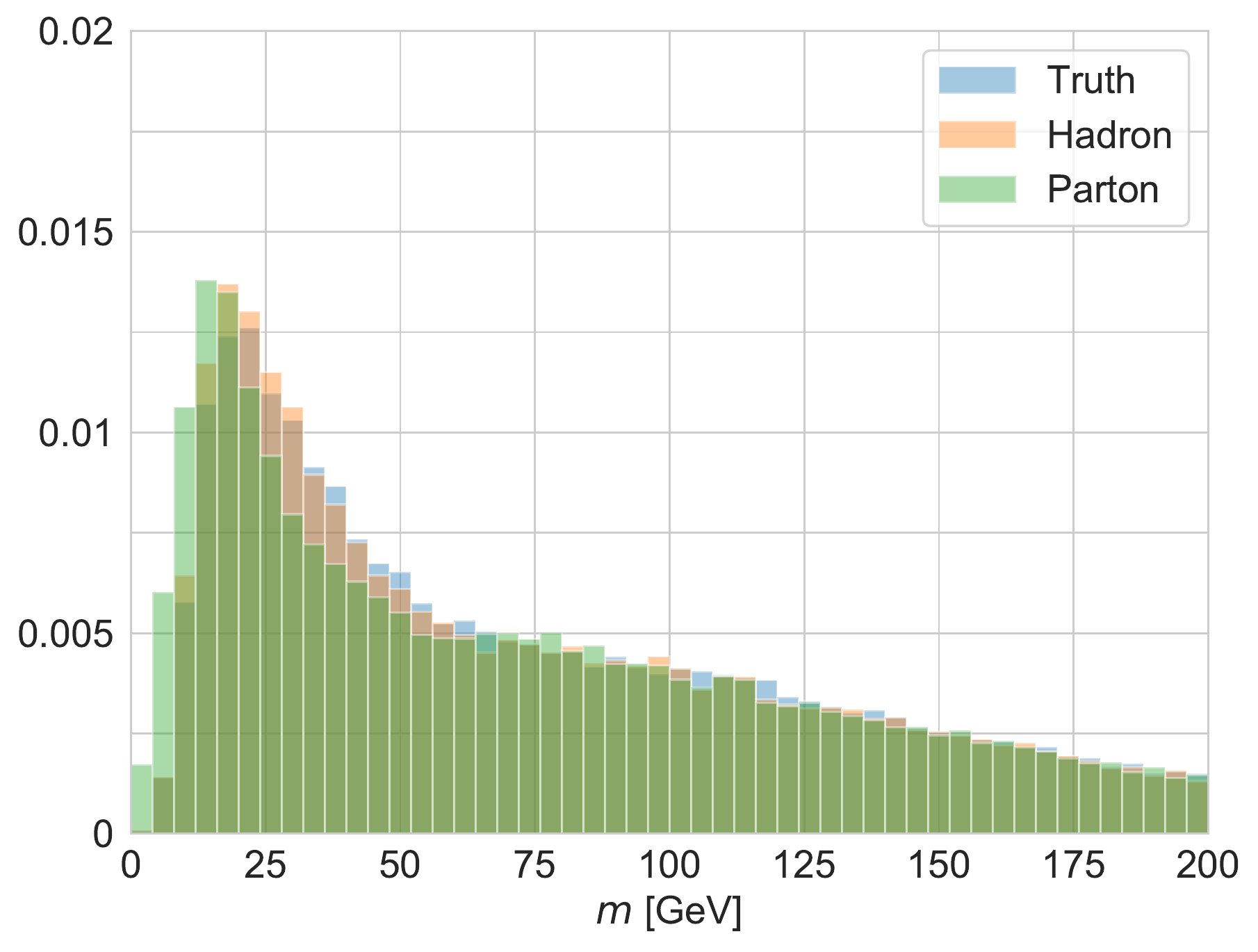}%
    \label{fig:np-effects-groomT}}~%
  \subfloat[RSD]{\includegraphics[width=0.45\textwidth]{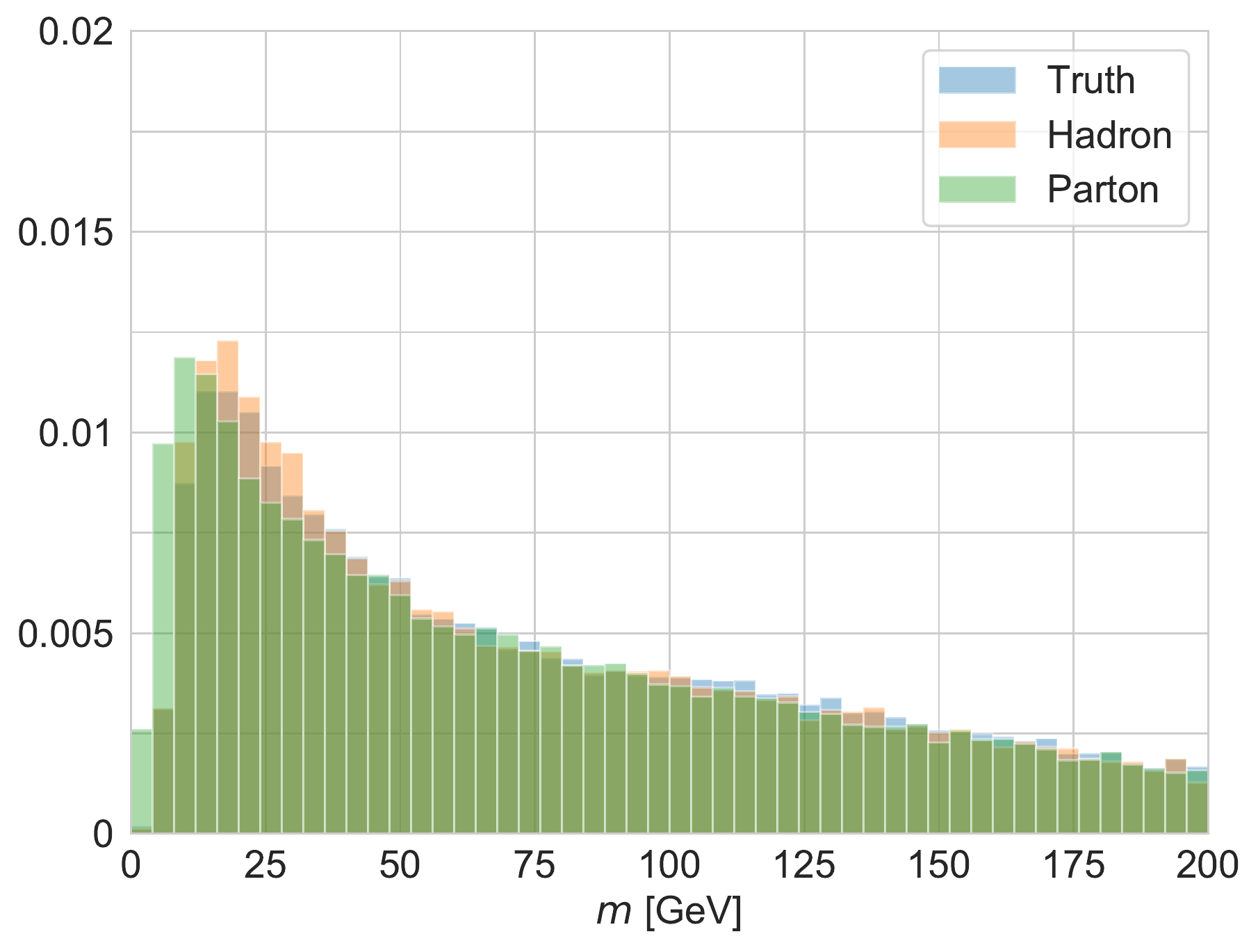}%
    \label{fig:np-effects-RSD}}%
  \caption{Jet mass spectrum for QCD jets at parton-level (green),
    hadron-level (orange) and including underlying event (blue).
    Distributions are shown for ungroomed jets (a), as well as
    after grooming with {\tt GroomRL} trained on $W$ data (b),
    on top data (c) or with RSD (d).}
  \label{fig:np-effects}
\end{figure*}

The jet mass spectrum for each sample is shown in
figure~\ref{fig:np-effects}, with results for ungroomed jets as well
as after grooming with \texttt{GroomRL-W}, \texttt{GroomRL-Top} and
RSD.
One can see immediately that the ungroomed jet mass spectrum is
strongly affected by non-perturbative effects, while groomed jets
become much more robust to these contributions.
For masses $m>50$ GeV, both \texttt{GroomRL} models become very
robust, showing a resilience to hadronization and underlying event
similar to that of RSD.
In the low mass range, \texttt{GroomRL} remains robust to multiple
parton interactions, but starts to show some dependence on
hadronization effects.

We note that no parton-level or hadron-level data was used in the
training, such that one would not a priori expect the derived
algorithm to be particularly resilient to these effects.
Although \texttt{GroomRL} already performs surprisingly well, one
could easily further improve the robustness of the model by including
some of this data with a suitable modification of the reward function
in the training of the DQN agent.

\section{Conclusions}

We have shown a promising application of RL to the issue of jet
grooming.
Using a carefully designed reward function, we have constructed a
groomer from a dense NN trained with a DQN agent.

This grooming algorithm was then applied to a range of data samples,
showing excellent results for the mass resolution of boosted heavy
particles.
In particular, while the training of the NN is performed on samples
consisting of $W$ (or top) jets, the groomer yields noticeable gains
in the top (or $W$) case as well, on data outside of the training
range.

The improvements in resolution and background reduction compared to
alternative state-of-the-art methods provide an encouraging
demonstration of the relevance of machine learning for jet grooming.
In particular, we showed that it is possible for a RL agent to
extract the underlying physics of jet grooming and distill this
knowledge into an efficient algorithm.

Due to its simplicity, the model we developed also retains most of the
calculability of other existing methods such as Soft Drop.
Accurate numerical computations of groomed jet observables are
therefore achievable, allowing for the possibility of direct
comparisons with data.
Furthermore, given an appropriate sample, one could also attempt to
train the grooming strategy on real data, bypassing some of the
limitations due to the use of parton shower programs.

The {\tt GroomRL} framework, available
online~\cite{groomRL,groomRL_lib}, is generic and can easily be
extended to higher-dimensional inputs, for example to consider
multiple emissions per step or additional kinematic information.
While the method presented in this article was applied to a specific
problem in particle physics, we expect that with a suitable choice of
reward function, this framework is in principle also applicable to a
range of problems where a tree requires pruning.

\textbf{Acknowledgments:}
We are grateful to Jia-Jie Zhu and Gavin Salam for comments on the
manuscript and to Jesse Thaler for useful discussions.
We also acknowledge the NVIDIA Corporation for the donation of a Titan
Xp GPU used for this research.
F.D.\ is supported by the Science and Technology Facilities Council
(STFC) under grant ST/P000770/1. S.C.\ is supported by the European 
Research Council under the European Union's Horizon 2020 research and
innovation Programme (grant agreement number 740006).

\bibliography{groomRL}

\end{document}